Authors:
*Olga Viberg*
KTH Royal Institute of Technology
Lindstedsvägen 3, 10044 Stockholm, Sweden
oviberg@kth.se

*Åke Grönlund*
Örebro University School of Business, Sweden
Fakultetsgatan 1, 70182 Örebro
ake.gronlund@oru.se

# Chapter 1. Introducing Practicable Learning Analytics

**Abstract**
Learning analytics have been argued as a key enabler to improving student learning at scale. Yet, despite considerable efforts by the learning analytics community across the world over the past decade, the evidence to support that claim is hitherto scarce, as is the demand from educators to adopt it into their practice. We introduce the concept of *practicable learning analytics* to illuminate what learning analytics may look like from the perspective of practice, and how this practice can be incorporated in learning analytics designs so as to make them more attractive for practitioners. As a framework for systematic analysis of the practice in which learning analytics tools and methods are to be employed, we use the concept of Information Systems Artifact (ISA) which comprises three interrelated subsystems: the *informational*, the *social* and the *technological* artefacts. The ISA approach entails systemic thinking which is necessary for discussing data-driven decision making in the context of educational systems, practices, and situations. The ten chapters in this book are presented and reflected upon from the ISA perspective, clarifying that detailed attention to the *social artefact* is critical to the design of practicable learning analytics.

## 1.1   Introduction

This book is about *practicable learning analytics*. So, let us begin by defining what we mean by *learning analytics* and by *practicable*. Learning analytics has over the last ten years become an established field of inquiry and a growing community of researchers and practitioners (Lang et al. 2022). It has been suggested as one of the learning technologies and practices that will significantly impact the future of teaching and learning (Pelletier et al. 2021). It is argued to be able to improve learning practice by transforming the ways we support learning and teaching (Viberg et al. 2018)

Learning analytics has been defined in several ways (Draschler and Kalz 2016; Rubel and Jones 2016; Xing et al. 2015). A widely employed and accepted definition explains it as the "measurement, collection, analysis and reporting of data about learners and their contexts, for the purposes of understanding and optimizing learning and the environments in which it occurs" (Long and Siemens 2011, p.34).

In order to recognise the complex nature of the learning analytics field, its related opportunities and corresponding challenges, researchers have stressed a need to further define and clarify what "kinds of improvement [in education] we seek to make, the most productive paths towards them, and to start to generate compelling evidence of the positive changes possible through learning analytics" (Lang et al. 2022, p. 14). Such evidence has so far been scarce and, to the extent it exists, it is often limited in scale (e.g., Ferguson and Clow 2017; Ifenthaler et al. 2021; Gašević et al. 2022). What does exist is predominantly found in higher education settings (e.g., Viberg et al. 2018; Wong and Li 2020; Ifenthaler et al. 2021); in K-12 settings, learning analytics research efforts have hitherto been limited (see e.g., de Sousa et al. 2021). If learning analytics can deliver on its promises, K-12 is arguably an even more important practice to improve as it concerns many more students and is more critical to society as it serves to educate the whole population, which makes it an even more complex field of activity.

In all educational contexts, there is a need to deliver on the promises of learning analytics and translate the unrealised potential into practice for improved learning at scale. But clearly learning analytics cannot be simplistically "put into practice", it has to be adopted into practice by practitioners who see a need for it and practical ways of using it. It has to be practicable.

*Practicable* suggests that something is "able to be done" or "put into action" or practised "successfully" (Cambridge Dictionary 2022; Oxford Learner's Dictionary 2022). This raises some questions: What exactly is that 'something' in learning analytics? Who is going to put it into practice? What practices are learning analytics aiming to improve? and How can we distinguish between what is more or less practicable? Would not it be good to have a theory for that, rather than just focusing on different aspects of learning analytics examinations, such as self-regulated learning (e.g., Montgomery et al. 2019; Viberg et al. 2020), collaborative learning (e.g., Wise et al., 2021) or social learning (e.g., Kaliisa et al. 2022). While these diverse learning analytics efforts are both interesting and meaningful to support, it is worthwhile to look at learning and teaching in a more systemic way, looking beyond isolated activities and considering them as a whole system orchestrated for students learning Education is composed of many activities conducted by both students and teachers, and affected by environmental factors. The latter includes many factors ranging from physical, like light and noise in the classroom, to social, like class sizes and composition and attitudes to learning in the home. Changes in one of those activities or factors may affect the others and may hence have consequences for the learning outcomes. It is not necessarily the case that focusing specifically on improving one factor leads to overall improvement of the system as a whole.

For example, Zhu, analysing data from the Programme for International Student Assessment (PISA), showed that reading literacy was significantly more important than mathematics for achievements in science (Zhu 2021), it was also directly influential on their mathematics achievements. Similarly, in a quasi-experimental study, Agélii Genlott and Grönlund (2016) introduced an ICT-supported method for improving literacy training in primary school and found that not only students' literacy achievements but also those in mathematics improved significantly, as measured by the national standard tests.

Such findings suggest that there are complex relations involved in learning; if you want to improve students' skills in mathematics and science, improving literacy training may be a good way to go. It certainly appears to be a bad idea to reduce literacy training to increase the time spent on mathematics training. So let us consider education practices from a systemic perspective.

## 1.2  A Systemic perspective on education practices

Making the use of learning analytics come into use in everyday teaching and learning activities at scale requires the tools and methods use to fit with the educational environments in which they are to be used. However, educational systems and activities are manifold and diverse, and even a brief analysis shows a great variety of situations and undertakings, as well as several stakeholders who may have different interests in learning analytics.

*Stakeholders.* Students and teachers are the frequently focused stakeholders in the learning analytics literature (e.g., Draschler and Greller 2011; Gašević et al. 2022; Gray et al. 2022), but educational leaders and school administrations are also involved and, in particular for younger students, parents have interest and take some part. These stakeholders play different roles and do not necessarily share the same view of what should be done in an educational institution and how to do that. While teachers and students take the keenest interest in the actual learning and teaching activities, parents, institutional leaders and school administrations are typically more interested in the results, often in the form of grades. Stakeholders can also include educational technology companies (e.g., learning management systems providers) bringing a commercial interest, and also researchers acting in the field. In sum, there are many stakeholders who may have quite different needs and interests in learning analytics (e.g., Sun et al. 2019), and this needs to be carefully considered when planning any learning analytics undertaking. It is easy to see that several conflicts between the interests of different stakeholders may come up. For example, Wise, Sarmiento and Boothe (2021) note that student and teacher stakeholders often fear that learning analytics systems are less about improving education and more about serving surveillance needs of the administration. They use the concept of "subversive learning analytics" to discuss the need to take a critical stance in order to disclose hidden assumptions built into technology designs.

*Situations.* Teaching and learning situations are quite different in school (especially primary and secondary) than at the university. Furthermore, learning frequently takes place with no teacher present and outside of school or scheduled classes at the university. The amount of individual student work and the responsibility of students to study independently increases as students get older, but it is also influenced by the number of teachers available, goals of educational programs, pedagogical approaches as well as educational and cultural contexts. Different study subjects require or entail certain activities, which may involve practical operations, movement, communication, testing, group work, and more. Some involve learning specific concepts, some involve understanding of systems, structures, logical reasoning, causes and effects in physical, social, or psychological matters, or all of these in combination.

In an average week, a student meets several teachers, several topics, and several situations. But common for them all is that there is some *information* to be handled and this takes place in a *social context*. As for the information, it is not only a content, it also has a form. It is typically written, audio or visual, but it may also be haptic or even tacit, such as when for example social behavioural norms are communicated by actions or non-actions. In an educational context, information must be presented in a form that is conducive to learning.

Introducing new technology, such as a novel learning analytics system, into an educational setting means changing both the situations and the information, and one cannot be changed without changing the other. For example, changing from reading a textbook to listening to the teacher means you have to stop listening to music on your headphones. This means that technology can also be seen as an actor in the social situation as it affects the conditions for student learning in several ways: in some situations, leading to improved learning but in others resulting in negative learning outcomes. That is, we cannot expect any new learning analytics tool introduced in a selected educational context to influence student learning directly and positively (as anticipated by designers); it changes the conditions in which learning activities occur, but the actual effect depends both on the technology and the situation, and it can be positive or negative. Often it is both; some of the anticipated positive effects may occur but also some "unintended consequence" that may be negative. The better we understand the situation before we intervene, the more likely we will design technology that has positive effects and no, or minimal, negative ones.

For at least fifty years, the discipline of information systems has been concerned with the introduction of information technology into people's work situations, that is, changing the social and informational situation of work. Pioneering in this regard was the Tavistock Institute in London where the concept of sociotechnical systems was coined (Emery and Trist 1960). Sociotechnical systems analysis and design was developed in the field of information systems design in the 1970s and onwards, pioneered by the Manchester Business School where Enid Mumford was a portal figure in the field of information systems, for example by developing the human-centred systems design method ETHICS (Effective Technical and Human Implementation of Computer Systems) (Mumford and Weir 1979).

The sociotechnical approach has since seen many developments, many new models and methods for analysis and design. The areas of work affected by digitalisation of tools and processes have multiplied – and education is among the most recent to be explored, decades after office work. An increasing number of theories have also come to use for analysing the relations between people and technology – and between *people*, *organisations* and *technology*. As an example, Wise et al. (2021) discuss critical learning analysis, critical race theory, speculative design and – still going strong! – sociotechnical systems.

## 1.3 The "Information System Artefact" in learning analytics

The research field of Learning Analytics is situated in the intersection of Learning, Analytics and Human-Centred Design (SOLAR 2021). "Learning" includes (at least) educational research, learning and assessment sciences, educational technology, "analytics" comprises, e.g., statistics, visualisation, computer/data sciences, artificial intelligence (but also qualitative analyses, such as critical analysis), and "human-centred design" is concerned with issues like usability, participatory design, sociotechnical systems thinking (SOLAR 2021). All these aspects are critical to successful implementation of learning analytics and require a carefully considered, approach to not only measure, but to better explain the targeted learning or teaching activities or processes

The disciplines of informatics (often named information systems) and computer science both share the interest in information technology artefacts, but informatics is distinguished by its focus on the user, which is in line with recent efforts on *human-centred* learning analytics (e.g., Buckingham Shum et al. 2019; Ochoa and Wise 2021). Who are the users of these technologies? What do they do? and How can technology help them do better? The object of study is people and technology *together*, and the concept of "information system" is typically defined as "a formal, sociotechnical, organizational system designed to collect, process, store, and distribute information" (Piccoli et al. 2018, p. 28).

A theoretical expression of that interest in users and use contexts is the notion of the *Information System Artefact* (ISA), as distinct from the information technology artefact (Lee et al. 2015). The ISA is "a system, itself comprising of three subsystems that are (1) a *technology* artefact, (2) an *information* artefact and (3) a *social* artefact, where the whole (the ISA) is greater than the sum of its parts (the three constituent artefacts as subsystems), where the information technology artefact (if one exists at all) does not necessarily predominate in considerations of design and where the ISA itself is something that people create" (i.e. an 'artefact'; Lee et al. 2015, p.6). The three sub-artefacts are interrelated and interdependent, which means that 'improving' one of the artefacts (in the literature, typically the technical, e.g., a learning analytics service) may in fact lead to a deterioration of the ISA. What is considered an improvement in any subsystem is only that which contributes to improving the whole, the ISA.

To make a LA system '*practicable*' in our terms means understanding how it enhances the ISA as a whole in the targeted educational setting. The ISA should be understood as an object to be designed. Creating and implementing a learning analytics system means designing a technical, a social and an information artefact in such a way that they interact well to improve the overall ISA, ultimately leading to student improved learning. This argument echoes the earlier call for a more systemic approach to learning analytics (Ferguson et al. 2014; Gašević et al. 2019).

Lee et al. (2015) define the components of the ISA, the three sub-artefacts, in the following way:

The *technology artefact*: "a human-created tool whose raison d'être is to be used to solve a problem, achieve a goal or serve a purpose that is human defined, human perceived or human felt" (p. 8). In the learning analytics setting, it could be different tools such as learning dashboards (see e.g., Susnjak et al. 2022) or other tools aimed at, for example, supporting students' self-regulated learning (for overview, see Perez-Alvarez et al. 2022) or formative feedback on academic writing (e.g., Knight et al. 2020) or collaborative peer feedback (e.g., Er et al. 2021).

The *information artefact*: "an instantiation of information, where the instantiation occurs through a human act either directly (as could happen through a person's verbal or written statement of a fact) or indirectly (as could happen through a person's running of a computer program to produce a quarterly report)" (p.8). The role of the *information* artefact in an educational setting can be to "form meaning", i.e., learn something, but it can also be other things, such as process information (like a calculator) or serve as a structure for information exchange (e.g., the alphabet).

The *information artefact*, hence, includes all the information that is present in a learning situation (in the case of learning analytics). Some of this information is subject to learning (the subject content), some is contextual (e.g., what concerns work methods). Introducing a technology artefact in an existing learning situation changes the information artefact insomuch as some new information may be added and some already existing information may appear in a different form (e.g., digital instead of physical or presented in a different digital format) or become available to students by different methods. This means any new learning analytics tool (a technology artefact) will in some way affect the information artefact of an educational context.

The *social artefact* "consists of, or incorporates, relationships or interactions between or among individuals through which an individual attempts to solve one of his or her problems, achieve one of his or her goals or serve one of his or her purposes" (p. 9). *Social* here means not just specific situations, like when a number of people meet and communicate, but also established, persistent relations such as institutions, roles, cultures, laws, policies and kinship.

In a simple way, the social artefact can be thought of as 'the classroom'. In a physical classroom, there are people with relations: professional and social. Professional relations concern the formal and technical part of teacher-student interaction (the teaching and learning activities), which is partly a function of the way it is organised as concerns, rules of conduct, time allocation, physical environment, class size, examination forms, and more. Social relations concern students' relations to each other, but also students' relation to schoolwork – which differ from very positive and uncomplicated to very negative and complicated – and the nature of the student-teacher communication, which is very much dependent on the personalities of the people involved.

The *social artefact* is much affected by changes in both the technology and the information ones. For example, when a new technology artefact is introduced in the classroom (the social artefact), it may mean that information that previously was physically available (e.g., a paper textbook or a teacher writing on a whiteboard) becomes part of the *technology* artefact and accessed and manipulable in new ways, the teacher-student communication changes. Teachers may have to spend time explaining to students how to handle the new tool, or students have to explain to teachers how they use them. Teachers may be less able to inspect students' work as it no longer is visible in the same way as previously when they could overview the work of an entire class in a moment. A 'social inspection' available by physical means – looking around in the classroom and then observing both individual work and social contacts – is to some extent replaced by an individual one available only through technical means (to the extent that the learning analytics application allows for that). Taken together, this means a change in the *social artefact* reducing the amount of physical communication and increasing the amount of technology-mediated communication. To what extent the quality of the social artefact is increased or reduced is subject to analysis, which is often not straight-forward.

Using the ISA model, different stakeholders' views of, and relation to, learning analytics systems, the information they use and produce, and the role they could play in teaching and learning environments can be more clearly identified and analysed. Teaching and learning are complex phenomena taking place in (different) social contexts, and the ISA model provides an analytical framework that includes those contexts.

## 1.4     Overview of the chapters

This book includes ten chapters (except this introductory chapter) that illustrate the examples and aspects of the *practicable* learning analytics efforts and related opportunities and challenges across three continents. Most concern higher education contexts. Whereas the first five chapters explicitly demonstrate institutional efforts to put learning analytics into practice at scale, the other five illustrate relevant efforts focusing on various aspects that are important to putting learning analytics into teaching and learning practice effectively.

In *Chapter 2,* Buckingham Shum (2023) presents and critically reflects on the efforts of an Australian public university to design, pilot and evaluate learning analytics tools over the last decade. These efforts are summarised as conversations in the *Boardroom*, the *Staff Room*, the *Server Room* and the *Classroom,* reflecting the different levels of influence, partnership and adaptation necessary to introduce and sustain novel technologies in the complex system that constitutes any educational institution.

In *Chapter 3*, Rienties et al. (2023) demonstrate how the (UK) Open University's Learning Design Initiative (OULDI) has been adopted and refined in a range of institutions to fit local and specific needs across three European projects, involving practitioners from nine countries. This chapter stresses that applying and translating the OULDI and learning analytics in other institutions and borders "is not a merely copy-paste job" since it requires a number of adaptations at different implementation levels, highlighting the importance of considering the targeted context. These required adaptations have been 'translated' into and presented as the Balanced Design Planning approach in the context of The University of Zagreb (Croatia).

In *Chapter 4,* De Laet (2023) illustrates two cases of learning analytics implementations at the institutional level in the context of Belgian higher education. The first case reflects an institutional path of bringing learning analytics to advising practice, and the second one presents the ongoing institutional efforts of bringing predictive analytics to advising practice, an approach building on explainable artificial intelligence to uncover the existing black-box predictions.

*Chapter 5* presents a project of "Learning Analytics – Students in Focus" in the context of another European university, TU Graz University of

Technology. Through the lens of the human-centred learning analytics approach, Barreiros et al. (2023) illustrate the iterative design, analysis, implementation and evaluation processes of the three learning analytics tools (i.e., the planner, the activity graph, and the learning diary), all contained in the student-facing dashboard.

In *Chapter 6,* Hilliger and Pérez-Sanagustín (2023) introduce the LALA CANVAS: a conceptual model to support a participatory approach to learning analytics adoption in higher education. The model has been employed across four Latin American universities affiliated with the LALA (Building Capacity to Use Learning Analytics to Improve Higher Education in Latin America) project. The LALA CANVAS model is argued to be a useful model to formulate change strategies in higher education settings where the adoption of learning analytics is still at an early stage.

In *Chapter 7,* Järvelä et al. (2023) present their recent empirical progress on metacognitive awareness and participation in cognitive and socio-emotional interaction to support the adaptive collaborative learning process. In particular, the authors present how learning process data and multimodal learning analytics can be used to uncover the regulation in computer-supported collaborative learning settings. They also provide a set of practical implications to assist students in collaborative learning activities.

In *Chapter 8*, Kizilcec and Davis (2023) introduce the current state of learning analytics education across the globe. This chapter contributes to practicable learning analytics by providing evidence on the status quo of teaching and learning analytics with a comprehensive review of current learning analytics programs, topics and pedagogies focused. This is followed by an in-depth case study of a learning analytics course offered to the students at Cornell University. Finally, a set of actionable guidelines for the community to consider when designing learning analytics courses is offered.

In *Chapter 9,* Glassey and Bälter (2023) present novel student data that learningsourcing produced. The aim is to marry learnersourcing efforts with learning analytics in terms of the types of novel learning data that is produced. The chapter provides a background to the emergence of learnersourcing as a topic, a taxonomy of the types of learnersourcing data and their supporting systems that increasingly make learnersourcing practicable for learning analytics. They also discuss challenges for using such data for learning analytics, for example as concerns data quality.

In *Chapter 10,* Viberg et al. (2023) argue for the importance of addressing cultural values when designing and implementing learning analytics services across countries. Viewing culture from a value-sensitive perspective, this chapter exemplifies two selected values (privacy and autonomy) that might play an important role in the design of learning analytics systems and discusses opportunities for culture- and value-sensitive design methods that can guide the design of culturally aware learning analytics systems. A set of design implications for culturally aware and value-sensitive learning analytics services is offered at the end.

Finally, in *Chapter 11,* Mavroudi (2023) reflects on the challenges associated with the ethical use of learning analytics in higher education, and how different selected policy frameworks address these challenges. It concludes with a list of practical recommendations on how to counteract specific challenges that might originate in the nature of learning analytics.

## 1.5   The chapters in context

Looking at the chapters in the book from the perspective of the ISA model, we find that most of them concern changes in the social artefact. In plain words that means changes in the way education is conducted. Education is somebody's work – teachers and students. Changing somebody's work from the outside – such as when introducing a learning analytics tool or system – will inevitably meet resistance unless it is clear to the people working education that there is not something negative in it for them. The starting point is often a suspicion that there is – most professions tend to believe that they are the ones who best understand how to do their job, so if someone from outside demands a change professionals tend to suspect that there is another agenda at play.

For changes to be positively received, there should also be something *positive* in it for them. Even if positive effects for teachers and students can be expected they can be very hard to argue in a convincing way as they may be difficult to measure and as they often appear later while there is always more work upfront when new systems are introduced.

The changes presented in the chapters in this book always concern the *social artefact*, changes in teachers' and students' daily work environment. Sometimes those changes are effects of changes in the other artefacts, the *technical* or the *information*al. Other times, changes in the social artefact motivates changes in one or both of the others. In all cases, changes in one artefact entails changes in another, and these changes are not always foreseen or planned for. In plain words, intended changes often lead to unintended consequences.

In the highly pragmatic *Chapter 2,* Buckingham Shum (2023) describes the entire setting in which learning analytics is to be implemented in terms of different "rooms". These rooms, which contain different stakeholders correspond quite directly to the different artefacts within ISA, and the chapter clearly describes the differences, and potential conflicts, between the different rooms. The *Staff Room*, the *Classroom,* and the *Boardroom* concern the *social* and *information* artefacts and focus on the required engagement of the different stakeholders involved: the university senior leadership, tutors, academics, students and teachers with learning analytics. But the *social* artefacts in the different rooms are different, representing different stakeholders' views and needs. In the *Staff Room* and the *Classroom*, there are teachers who are engaged in engaging with students and their work, and with the knowledge content of their courses, and who want to have information that can help them with that. In the *Boardroom*, university leadership is working in a business environment where the interest is in information about performance on university strategic priorities and how to improve return on investment in production. The learning analytics entrepreneur must engage both these audiences, but the way to do it differs as each room has different requirements on the information artefact. Information about teaching and learning, pedagogical issues and students' learning processes, is of interest for teachers, but in the Boardroom, there are rather requirements for information about production costs and results, including, for example, process effectiveness and efficiency, and performance of teachers. Not only is such information not interesting to teachers, it may even be discouraging to find that their own performance is monitored through the new system. The *Server Room* concerns the engagement with the information technology services, that is, the *technology artefact,* which is also critical to the success of any learning analytics implementation. Here, one important interest is how a new learning analytics system fits in with the existing ecosystem of applications which it needs to be able to interact with. This is not just a technical issue, the degree of integration among technical systems directly affects students' and teachers' work in the classroom.

*Chapter 3* in presenting the new approach to learning analytics to fit local institutional needs across several European institutions, stresses the importance of the *information* and *social* artefacts but also the situational nature of them. Both the information handling, and the social setup in which the system was to be used, were areas where most adaptations to the system had to be made to fit the way education and administration were organised and conducted in different places due to regulation and practices, and at both national and local (university) level. These regulations were implemented in work instructions and practices of administrators, managers, and professionals, and in technical systems, which together formed a very firm social infrastructure to which any new work process must adapt. While there was less adaptation needed for the *technology* artefact in the case presented, this, too, needed concern as there has to be a sufficiently mature technical infrastructure in an organisation to be able to implement any learning analytics system.

*Chapter 4* reflects on the interrelations between the three ISA sub-artefacts when presenting the scale-up process of the advising dashboard. The impetus to change came from an improved *technology* artefact aimed to improve the *information* artefact, that is, lead to better information handling and hence more effective work processes, specifically by supporting the dialogue between academic advisors and students. The change process involved several challenges related to the *social artefact*, including "overcoming resistance to change, alignment with educational values of the higher education institute, and tailoring to the particular context". Interestingly – and in contrast to similar efforts previously reported in the literature, the project was successful in terms of improving the *social artefact* – it resulted in the academic advisors (the key system users) feeling that the system made them better equipped to conduct a constructive and "more personal" dialogue with students. The author attributes this success to two main factors. First, the system did not include any prescriptive or predictive components, which are often found to be sources to resistance because they interfere uninvitedly in people's work (negatively affect the social artefact, in terms of ISA). Second, the implementation project took a bottom-up approach with the goal of supporting the advising dialogue and the professionals were included in an iterative user-centred design process, hence giving them an element of ownership and control of the new system.

*Chapter 5* discusses a human-centred approach to LA design, which means the point of departure is the *social artefact*; the aim of a human-centred approach is to design work processes, work organisation, and technical systems to fit people. The chapter describes a project where use cases were first constructed. This was done by defining students' personas and descriptions of several scenarios illustrating when and with what intent the students may use the learning analytics dashboard to acquire or develop self-regulated skills, and how they might act to achieve a goal using the dashboard. Based on a selection of these scenarios, the project went on to produce design solutions, which were then moved forward to prototypes for testing with the intended users. In terms of the ISA, this means designing the entire ISA artefact using the *social artefact* (the scenarios) as the reference and as a test for the quality of the other two sub-artefacts. The prototypes represent the information and the technology artefacts. They were based on the scenarios; the *information artefact* concerned selecting which information to include and how to organise it to meet user needs, and the *technology artefact* concerned implementing the user interface to that information in such a way that it provides adequate support to their use processes. This shows a mutual dependence among the sub-artefacts. The social artefact informed the design of the information and technology artefacts, but the latter two also informed the design of the social artefact; during the design process, the prototypes were used to make the scenarios more concrete to users.

Similar to the previous chapter, *Chapter 6* also starts from an interest in the *social artefact.* The contribution here is a conceptual model to support a participatory approach to learning analytics adoption in higher education; that is, a way to understand the social environment in which learning analytics is to work by means of direct user participation. The challenge is to discuss learning analytics at an early stage of development, which means it is still a rather hypothetical concept to participants as there is little in the form of examples of proven practice to guide prospective users' expectations. The method for discussion is group discussions, and the aim is to understand what needs there might be in educational practice that learning analytics could draw upon so as to be useful to practitioners. The model proposed and tested is built on factors known to be important for successful implementation: political context, influential actors, desired behaviours, internal capabilities, change strategy, and indicators and instruments for assessment and evaluations.

In *Chapter 7,* again the *social artefact* is in focus, this time in a basic research perspective. The study studies group collaboration with the aim to be able to support its regulation. Effective collaborative learning requires group members to ensure that they work toward the shared goals and in order to be able to regulate their work they need to reveal to each other when they become aware that their collaboration is not heading toward the shared goals. This regulation takes place not only by using words but also by social, visual, cues of different kinds. The research studies multimodal data from group processes to identify "socially shared regulation episodes" (Järvelä et al., 2023, p.X).

*Chapter 8* notes that higher education in learning analytics is conducted in different schools including not only Education but also Computer Science, Information Science and Media Studies. This means that both students and teachers come from a wide variety of disciplinary backgrounds, and many will not have a background in educational environments. The authors caution against overly focusing on numbers and – in the spirit of ISA, if not in the words – encourages educators to not forget the educational (social) environments where learning analytics are to be used (that is, the social artefact): "Before students are asked to conduct any analyses or learn a new programming language for data processing, it is critical that they first develop a strong foundational understanding of the field" (Kizilcec and Davis 2023, p.X). This understanding will help them select what (educational) problems to engage with.

*Chapter 9* concerns "learnersourcing", where the basic idea is to have students do part of the grading or each other's work by means of a (teacher-organised) peer-review process. This constitutes a major change in the way education is set up, that is the *social artefact*. It means the students must, to some part, assume a role as evaluator, which is quite contrary to the traditional role where they (individually or in cooperation) submit work for evaluation to another stakeholder in the setting, the teachers. It also means the teachers back off a little from the evaluation process by delegating parts of it to students. The main driver behind the change is to save teachers' time by letting students do some of the information processing required for assessment of student work. In terms of the ISA, this means rearranging the *information artefacts*, and this change has considerable effects on the social artefact. This change redistributes some workload/information processing, but also changes the roles of stakeholders. It forces students to view their assignments from the perspective of teachers and stated quality criteria. It also changes the role of a teacher who becomes less of a direct actor and more of a "learning manager" overviewing a learning system (of students working in a digital tool) and intervening only as necessary.

*Chapter 10* discusses how cultural values can be critical to learning analytics use, and how to make learning analytics design and related examinations "culturally aware" and "value-sensitive". While culture is a concept that eludes strict definition, there are several cultural values that may strongly influence the social environment (the *social artefact*) that can be more clearly defined and that are valued differently in different countries. The chapter discusses two out of a set of such culturally significant values – privacy and autonomy – and discusses how design methods can take values into consideration.

*Chapter 11* seeks to contribute to the discussion on the ethical usage (e.g., as concerns transparency, privacy, access) of learning analytics in higher education by examining the main theoretical concepts in the field against respective policies or codes of LA ethics at several selected universities in three countries.

This discussion directly concerns the *information* and the *technology artefacts* (how data about individuals is handled in a digital environment) but it more fundamentally concerns the *social artefact* as ethics is basically a social contract. The key to using data on individuals is consent by the individuals themselves. The legal regulation provides a – very strict – framework, but as many situations require data that is more or less personal and sensitive, consent is the method used to be able to retrieve and manipulate such data. In online shopping and social media, explicit consent is needed – "I agree to allow cookies" – but in education, there is a social contract between teachers and students that teachers can use some student data for the purpose of being able to teach them. Some of that data may be sensitive, like students' medical diagnoses and other personal characteristics, personal background, and views, which may affect learning and require special teaching methods. The condition to use such data is discretion; it is only for use in teaching situations. This condition is typically implicit, it is not expressed in personal social contracts but comes with the definitions and practices of the educational environments and professions (that is, social contracts at national level). Hence, it differs across countries. Physical educational environments make it easy to meet the contract terms, as each teacher is in control of the data.

LA changes this as much data that may be sensitive is handled digitally, and the ways that this is done is not only beyond the control of teachers and students, but also often opaque and difficult for them to learn about.

This means that the policies of educational institutions become important. This chapter discusses higher education, but the issues discussed are arguably even more important to K-12 education as it concerns more students, younger students (and therefore also involves their parents) and generally a more diverse population.

## 1.6 Conclusion

The chapters in this book together bring up many issues pertinent to making learning analytics more *practicable.* They all focus on specific issues or practices and use different theoretical perspectives but for the purpose of discussing the overall perspective of 'practicability', we have provided an overview of the problem of making learning analytics practicable by using the concept of the *Information System Artefact* (ISA). The ISA consists of three integrated and mutually dependent sub-artefacts, *social, technical,* and *informational*. In the brief analysis of the chapters in the previous section, we provide glimpses of how the three sub–artefacts relate to each other in the different educational situations or aspects of learning that each chapter discusses. Throughout the chapters, it is clear that the *social artefact* is the most fundamental for practicability. Any substantive changes in information handling – content, process, format, technology used – will affect the social educational situations, and to be effective – or at all used – they will have to be understood and accepted by the practitioners involved. This is not to say that the *social artefact* – the way in which education is conducted – cannot or should not change. Quite to the contrary, practitioners – students as well as teachers – experience many problems or deficiencies in the way education is conducted and are likely to welcome changes, just like they have already done as concerns use of various other technologies. But the welcoming is contingent on them anticipating, and ultimately experiencing, benefits to their teaching and students learning. Therefore, an important key to successful large-scale implementation of learning analytics is the way teachers and students are approached. What is not practicable is not likely to be used.

### Acknowledgements

Many thanks to Simon Buckingham Shum who offered constructive feedback on the draft of this chapter, and all the authors and revie-wers who have contributed to this book.

**References**

Agélli Genlott, A & Grönlund, Å 2016, 'Closing the gaps – Improving literacy and mathematics by ict-enhanced collaboration', *Computers & Education,* vol.99, pp. 68-80.

Barreiros, C, Leitner, P, Ebner, M, Veas, E & Lindstaedt, S 2023, 'Students in focus – Moving towards human-centered learning analytics', In Viberg O & Gronlund, A (2023). *Practicable Learning Analytics.* Springer Nature, pp.xx

Buckingham Shum, S 2023, 'Embedding Learning Analytics in a University: Boardroom, Staff Room, Server Room, Classroom', In Viberg O & Gronlund, A (2023). *Practicable Learning Analytics.* Springer Nature, pp.xx

Buckingham Shum, S, Ferguson, R & Martinez-Maldonado, R 2019, 'Human-centered learning analytics', *Journal of Learning Analytics,* vol. 6, no.2, pp.1-9.

Cambridge Dictionary 2022 (online). https://dictionary.cambridge.org/

De Laet T (2023), 'Learning dashboards for academic advising in practice', In Viberg O & Gronlund, A (2023). *Practicable Learning Analytics.* Springer Nature, pp.xx

De Sousa, E, Alexandre, B, Ferreira Mello, R, Fontual Falcao, T, Vesin, B & Gasevic, D 2021, 'Applications of learning analytics in high schools: A systematic literature review, *Frontiers in Artificial Intelligence'*, vol. 4, no. 737891, pp.1-14.

Draschler, H & Greller, W 2012, 'The pulse of learning analytics understandings and expectations from the stakeholders', In LAK12: *Proceedings of the 2$^{nd}$ International Conference on Learning Analytics and Knowledge*, pp. 120-129.

Draschler, H & Kalz M 2016, 'The MOOC and learning analytics innovative cycle (MOLAC): A reflective summary of ongoing research and its challenges', *Journal of Computer Assisted Learning,* vol.32, no.3, pp. 281-290.

Emery, F E & Trist, E 1960, 'Socio-technical Systems', In Churchman C W M & Verhulst M (Eds.), *Management Sciences Models and Techniques.* Vol. 2, Pergamon Press, pp. 83–97.

Er, E, Dimitriadis, Y & Gasevic D 2021, 'Collaborative peer feedback and learning analytics: theory-oriented design for supporting class-wide interventions', *Assessment & Evaluation in Higher Education,* vol. 46, no. 2., 169-190.

Ferguson, R, Macfaydyen, L, Clow, D, Tynan, B, Alexander, S & Dawson, S 2014, 'Setting learning analytics in context: Overcoming the barriers to large-scale adoption', *Journal of Learning Analytics,* vol.1, no. 3, pp.120-144.

Gašević, D, Tsai, Y-S, Dawson, S & Pardo, A 2019, 'How do we start? An approach to learning analytics adoption in higher education', *International Journal of Information and Learning Technology,* vol. 36, no. 4, pp. 342-353.

Glassey, R & Bälter, O 2023, 'Learningsourcing analytics', In Viberg O & Gronlund, A (2023). *Practicable Learning Analytics.* Springer Nature, pp.xx

Gašević, D, Tsai, Y.-S., & Draschler, H 2022. 'Learning analytics in higher education – Stakeholders, strategy and use', *Internet and Higher Education,* vol. 52, no. 100833, pp.1-5.

Gray, G, Schalk, A E, Cooke, G, Rooney, P & O'Rourke, K 2022, 'Stakeholders' insights on learning analytics: Perspectives of students and staff', *Computers & Education,* vol.187, no.104550, pp.1-16.

Hilliger I & Perez Sanagustín, M 2023, 'LALA Canvas: A model for guiding group discussions in early stages of learning analytics adoption', In Viberg O & Gronlund, A (2023). *Practicable Learning Analytics.* Springer Nature, pp.xx

Ifenthaler, D, Gibson, D, Prasse, D, Shimada, A & Yamada M 2021, *'*Putting learning back into learning analytics: actions for policy makers, researchers, and practitioners', *Educational Technology Research and Development,* vol.69, pp. 2131–2150.

Järvelä, S, Vuorenmaa, Cini, A, Malmberg, J & Järvenoja, H 2023, 'How learning process data can inform regulation in collaborative learning practice', In Viberg O & Gronlund, A (2023). *Practicable Learning Analytics.* Springer Nature, pp.xx

Kaliisa, R, Rienties, B, Morch, A & Kluge, A 2022. 'Social learning analytics in computer-supported collaborative learning environments: A systematic review of empirical studies', *Computers and Education Open,* vol. 3, no.100073, pp.1-11.

Kizilcec, R & Cohen, G 2017, 'Eight-minute self-regulation intervention raises educational attainment at scale in individualist but not collectivist cultures', In *Proceedings of the National Academy of Sciences*, vol.114, no.17, 4348-4353.

Kizilcec, R & Davis, D 2023, 'Learning Analytics Education: A case study, review of current programs, and recommendations for instructors', In Viberg O & Gronlund, A (2023). *Practicable Learning Analytics.* Springer Nature, pp.xx

Knight, S, Vijay Mogarkar, R, Liu, M, Kitto, K, Sándor, Á, Lucas, C, Wight, R, Sutton, N, Ryan, P, Gibson, A, Abel, S, Shibani, A & Buckingham Shum, S 2020, 'AcaWriter: A learning analytics tool for formative feedback on academic writing', *Journal of Writing Research*, vol.12, no.1, pp.141–186.

Lang, C, Wise, A F, Merceron, A, Gasevic, D & Siemens, G 2022, 'What is learning analytics?', In Lang, C, Siemens, G, Wise, A F, Gasevic, D & Merceron, A (Eds.). The Handbook of Learning Analytics, pp.8-18.

Lee, A, Thomas, M & Baskerville, R 2015, 'Going back to basics in design science: from the information technology artifact to the information systems artifact', *Information Systems Journal,* vol.25, pp. 5–21.

Martinez-Maldonado, R, Elliot, D, Axisa, C, Power, T, Echeverria, C & Buckingham Shum, S 2022. 'Designing translucent learning analytics with teachers: an elicitation process', *Interactive Learning Environments,* vol. 30, no. 6, pp.1077-1091.

Mavroudi, A 2023, 'Challenges and recommendations on the ethical usage of learning analytics in higher education', In Viberg O & Gronlund, A (2023). *Practicable Learning Analytics.* Springer Nature, pp.xx

Montgomery, A, Mousavi, A, Carbonaro, M, Hayward, D & Dunn, D 2019, 'Using learning analytics to explore self-regulated learning in flipped blended learning music teacher education', *British Journal of Educational Technology*, vol. 50, no.1, pp.114-127.

Mumford, E & Weir, M W 1979. Computer Systems in Work Design: The ETHICS method: Effective Technical and Human Implementation of Computer System, John Wiley & Sons.

Ochoa, X & Wise, A F 2021, 'Supporting the shift to digital with student-centered learning analytics', *Educational Technology Research and Development,* vol. 69, pp.357-361.

Oxford English Dictionary: The definitive record of the English language 2022. https://www.oed.com/

Pelletier, K, Brown, M, Brooks, C, McCormack, M, Reeves, J et al. 2021 'EDUCAUSE Horizon Report, Teaching and Learning Edition'. https://library.educause.edu/-/media/files/library/2022/4/2022hrteachinglearning.pdf?la=en&hash=6F6B51DFF485A06DF6BDA8F88A0894EF9938D50B

Perez Alvarez, R, Jivet, I, Perez-Sanagustin, M, Scheffel, M & Verbert, K 2022, 'Tools designed to support self-regulated learning in online learning environments: A systematic review', *IEEE Transactions of Learning Technologies,* vol. 15, no.4, pp.508-522.

Piccoli, G & Pigni, F 2018. *Information systems for managers: with cases* (Edition 4.0 ed.). Prospect Press.

Rienties, B, Balaban, I, Divjak, B, Grabar, D, Svetec, B & Vonda, P (2023), 'Applying and translating learning design and analytics approaches across borders', In Viberg O & Gronlund, Å (2023). *Practicable Learning Analytics.* Springer Nature, pp.xx

Rizvi, S, Rienties, B, Rogaten, J & Kizilcec, R 2022, 'Beyond one-size fits-all in MOOCs: Variation in learning design and persistence of learning in different cultural and socioeconomic contexts', *Computers in Human Behavior,* vol. 126, no. 106973. https://doi.org/10.1016/j.chb.2021.106973

Rubel, A & Jones, K. 2016, 'Student privacy in learning analytics: An information ethics perspective', *The Information Society,* vol. 32, no.2, pp.143-159.

Sarmiento, J P & Wise, A F 2022, 'Participatory and co-design of learning analytics: An initial review of the literature', In LAK 22: *12[th] International Learning Analytics and Knowledge Conference*, pp. 535-541.

Simon, H. (1996) The Sciences of the Artificial. MIT Press, Cambridge, Massachusetts

Sun, K, Mhaidli, A, Watel, S, Brooks, C & Schaub, F 2019, 'It's my data! Tensions among stakeholders of a learning analytics dashboard', *In CHI'19: Proceedings of the 2019 CHI Conference on Human Factors in Computing Systems,* pp. 1-14. https://doi.org/10.1145/3290605.3300824

Viberg, O, Jivet, I & Scheffel, M 2023, 'Designing culturally aware learning analytics: A value sensitive approach', In Viberg O & Gronlund, A (2023). Practicable Learning Analytics. Springer Nature, pp.xx

Viberg, O, Hatakka, M, Bälter, O & Mavroudi, A 2018, 'The current landscape of learning analytics in higher education', *Computers in Human Behaviour,* vol.89, pp.98-110.

Viberg, O, Khalil, M & Baars, M 2020, 'Self-Regulated Learning and Learning Analytics in Online Learning Environments: A Review of Empirical Research', In Proceedings of the 10th International Conference on Learning Analytics and Knowledge (LAK20), pp.524-533.

Xing, W, Rui, G, Petakovic, E & Goggins, S 2015, 'Participation-based student final performance prediction model through interpretable Genetic Programming: Integrating learning analytics, educational data mining and theory', *Computers in Human Behaviour,* vol.47, pp.168-181.

Wise, A F, Knight, S & Buckingham Shum, S B 2021, 'Collaborative Learning Analytics', In: Cress, U., Rosé, C., Wise, A.F., Oshima, J. (Eds). International Handbook of Computer-Supported Collaborative Learning. Computer-Supported Collaborative Learning Series, vol 19. Springer, Cham. https://doi.org/10.1007/978-3-030-65291-3_23

Wise, A, Sarmiento, J P & Boothe, M 2021, 'Subversive learning analytics', In LAK21: 11th International Learning Analytics and Knowledge

Conference, pp. 639-645. https://doi.org/10.1145/3448139.3448210

Wong, B T & Li, K C 2020, 'A review of learning analytics interventions in higher education', *Journal of Computers in Education,* vo.7, no.1, pp.7-28.

Zhu, Y 2022, 'Reading matters more than mathematics in science learning: an analysis of the relationship between student achievement in reading, mathematics, and science', *International Journal of Science Education*, vol. 44, no. 1, pp 1-17.